\documentclass[11pt]{article}

\usepackage[final]{acl}

\usepackage{times}
\usepackage{latexsym}
\usepackage{amsmath} 
\usepackage{multirow, enumitem}
\usepackage{booktabs}
\usepackage[T1]{fontenc}

\usepackage[utf8]{inputenc}

\usepackage{microtype}

\usepackage{inconsolata}

\usepackage{graphicx}

\title{DELULU: Discriminative Embedding Learning Using Latent Units for Speaker-Aware Self-Trained Speech Foundational Model}

\author{
  Massa Baali, Rita Singh, Bhiksha Raj \\
  Carnegie Mellon University \\
  \texttt{mbaali@cs.cmu.edu}
}

\begin{document}
\maketitle
\begin{abstract}
Self-supervised speech models have achieved remarkable success on content-driven tasks, yet they remain limited in capturing speaker-discriminative features critical for verification, diarization, and profiling applications. We introduce \textsc{DELULU}, a speaker-aware self-trained foundational model that addresses this limitation by incorporating speaker-informed structure into pseudo-label generation. DELULU leverages frame-level embeddings from ReDimNet, a state-of-the-art speaker verification model, to guide k-means clustering during pre-training, introducing a speaker-discriminative inductive bias that aligns representation learning with speaker identity. DELULU significantly outperforms prior SSL models across a range of speaker-centric tasks, achieving up to \textbf{62\% relative improvement} in equal error rate (EER) for speaker verification and consistent gains on zero-shot profiling tasks including gender, age, accent, and speaker counting; notably surpassing even its teacher model on zero-shot evaluations. Our findings demonstrate that \textbf{DELULU is a strong universal encoder for speaker-aware speech processing}, enabling superior performance without task-specific fine-tuning.

\end{abstract}

\section{Introduction}
Speaker information is essential for a wide range of speech-related applications, including speaker verification, diarization, and personalized speech generation \cite{reynolds2000speaker,anguera2012speaker,casanova2022yourtts}. Despite the recent success of self-supervised learning (SSL) in speech representation \cite{baevski2020wav2vec,hsu2021hubert,chen2022wavlm}, existing models still struggle to capture speaker-specific characteristics effectively. The lack of robust speaker-aware representations poses a fundamental limitation for building systems that rely heavily on identity cues \cite{qian2022contentvec,zhang2023introducing}.
Although self-supervised models have achieved strong results across a variety of speech and audio tasks \cite{baevski2020wav2vec,hsu2021hubert,chen2022wavlm,waheed2024speech}, their performance on speaker-related applications remains limited. The key bottleneck lies in their reliance on pseudo-labels generated through acoustic-only clustering, which are insufficiently aligned with speaker-discriminative structure. In HuBERT \cite{hsu2021hubert}, the k-means clustering step relies on shallow acoustic features that prioritize phonetic similarity, often suppressing speaker-specific information such as voice quality, prosody, and speaking style. Although WavLM \cite{chen2022wavlm} introduces additional context modeling and denoising objectives, it inherits the same clustering mechanism and, as a result, continues to struggle with learning robust speaker representations.
We introduce DELULU, a speaker-aware self-trained foundational model that explicitly incorporates speaker-informed structure into the pretraining process. DELULU leverages ReDimNet~\cite{yakovlev2024reshape}, a state-of-the-art speaker verification network, to guide k-means clustering with frame-level speaker embeddings rather than purely acoustic features. While ReDimNet itself is trained in a supervised manner and provides indirect supervision through pseudo-label generation, DELULU remains fundamentally self-trained in the classical sense \cite{yarowsky1995unsupervised,amini2025self}: it learns from its own predictions on unlabeled data, with the teacher serving only to initialize the target structure. Crucially, the pseudo-label signals DELULU derives from ReDimNet are misaligned with ReDimNet's original training objectives—ReDimNet optimizes for utterance-level speaker discrimination, whereas DELULU's clustering operates at the frame level for masked prediction—making the direct influence of ReDimNet's supervision minimal. This design introduces a speaker-aware inductive bias into the learning pipeline while preserving the scalability and generality of self-training, enabling the model to capture speaker-relevant information more effectively without compromising general acoustic modeling. 
A stronger speaker-oriented foundation model should yield representations that better encode speaker identity, leading to improved performance on forensic and identity-centric applications both in zero-shot settings and after task-specific fine-tuning. We show that this holds true: DELULU outperforms existing SSL models across speaker verification, profiling (age, gender, accent), and speaker counting tasks, with particularly large gains in zero-shot evaluation. These improvements translate directly to downstream fine-tuning, confirming that introducing speaker-aware structure at the clustering stage produces more transferable representations. Beyond performance, DELULU illustrates a broader principle: self-trained speech models can be strengthened by integrating targeted external signals to guide pseudo-label formation, without requiring direct task-aligned supervision. This framework provides a scalable path toward speaker-aware foundation models and can be generalized to other architectures and domains beyond speaker modeling.
Our main contributions are as follows:
\begin{itemize}[leftmargin=*]
    \item We introduce DELULU, a speaker-aware self-trained speech model that addresses the fundamental limitation of speaker discriminability in existing SSL approaches by integrating external speaker-informed structure into the pseudo-label generation process through ReDimNet-guided k-means clustering.
    \item We achieve state-of-the-art performance across a wide suite of speaker-centric benchmarks, including speaker verification, profiling (age, gender, accent), and speaker counting, demonstrating up to 62\% relative improvement in EER.
    \item We establish DELULU as a strong universal encoder for speaker tasks, outperforming prior SSL models by a large margin in zero-shot settings and showing competitive results in upstream performance comparisons—notably surpassing even its teacher model, ReDimNet.
\end{itemize}

\section{Related Work}
\subsection{Self-Supervised Speech Representation Learning}
Self-supervised learning (SSL) has revolutionized speech processing by enabling models to learn rich representations from large-scale unlabeled audio data~\cite{baevski2020wav2vec, hsu2021hubert, chen2022wavlm}. These models typically employ pretext tasks such as contrastive predictive coding or masked language modeling to capture phonetic and acoustic structures without explicit labels.
Wav2vec 2.0~\cite{baevski2020wav2vec} introduced a contrastive loss over quantized latent representations, achieving strong performance on speech recognition benchmarks. HuBERT~\cite{hsu2021hubert} advanced this by using offline k-means clustering on MFCC features to generate pseudo-labels, focusing on discrete unit discovery that aligns well with phonetic content. WavLM~\cite{chen2022wavlm} further incorporated denoising objectives and utterance mixing to improve robustness to noise and overlapping speech, making it suitable for a broader range of tasks including speaker identification.
Despite these advances, standard SSL models often underperform on speaker-centric tasks due to their emphasis on content over speaker identity~\cite{chen2022wavlm}. Recent efforts have explored scaling these models to multilingual settings~\cite{pratap2024scaling} or enhancing them with multi-task learning~\cite{hu2024wavllm}, but speaker discriminability remains a challenge.
\subsection{Supervised Speaker Representation Learning}
Traditional speaker verification systems rely on the fundamental assumption that the human voice is a unique biometric trait~\cite{singh2025human}. This expected uniqueness underpins decades of research into supervised speaker modeling pipelines. The x-vector system~\cite{snyder2018x} uses time-delay neural networks (TDNNs) with statistical pooling to produce fixed-dimensional utterance embeddings, achieving robust performance on benchmarks like VoxCeleb~\cite{nagrani2017voxceleb}.
Advancements include ECAPA-TDNN~\cite{desplanques2020ecapa}, which incorporates channel and context-dependent attention to better capture speaker variability. More recently, ReDimNet~\cite{yakovlev2024reshape} introduced a reshaped dimensionality network that optimizes for both local and global speaker features, setting new state-of-the-art results on speaker verification tasks.
However, these supervised embeddings are often over-specialized for identity discrimination, limiting their generalization to broader speaker profiling tasks such as age and gender estimation or forensic analysis~\cite{baalipdaf}. For instance, phonetic biases in attention mechanisms can confound speaker-specific cues with content-related artifacts, reducing transferability to non-identity attributes~\cite{baalipdaf}. These supervised approaches excel in controlled settings but require large labeled datasets and struggle with domain shifts~\cite{desplanques2020ecapa}. Integrating their strengths into SSL pipelines offers a promising direction to enhance unsupervised representation learning, enabling foundational models that produce versatile, speaker-aware representations suitable for diverse downstream applications.
\subsection{Self-Supervised Learning for Speaker Tasks}
A growing body of work adapts SSL specifically for speaker-related tasks such as verification and diarization, often leveraging contrastive or distillation-based objectives to learn speaker-discriminative representations from unlabeled data~\cite{lepage2025self}.
Early efforts applied contrastive learning frameworks from computer vision, such as SimCLR adapted for speech~\cite{jiang2020speech}, which minimizes distances between augmented views of the same utterance while maximizing separations from others.
Self-distillation approaches, inspired by DINO~\cite{caron2021emerging,ashihara2024self}, have shown particular promise. 
More recently, CoLMbo~\cite{baali2025colmbo} proposed a speaker language model that integrates prompt-conditioned speaker encoders to generate descriptive speaker profiles. This approach showed strong zero-shot generalization on demographic traits such as dialect and age, demonstrating the potential of using large-scale speaker models beyond classification tasks. Meanwhile, CAARMA~\cite{baali2025caarma} introduced an augmentation framework where HuBERT’s hidden layers serve as discriminators during adversarial training, improving the speaker discriminability of the learned embeddings. These methods suggest that integrating architectural and training-level bias can enhance SSL’s ability to encode speaker traits but they do not intervene directly at the pseudo-label generation step as DELULU does.

\textsc{DELULU} builds on these by directly guiding SSL clustering with supervised speaker features, bridging the gap between general acoustic modeling and targeted speaker discrimination.

\section{Architecture}
\begin{figure}[t]
  \centering
  \includegraphics[width=0.99\linewidth]{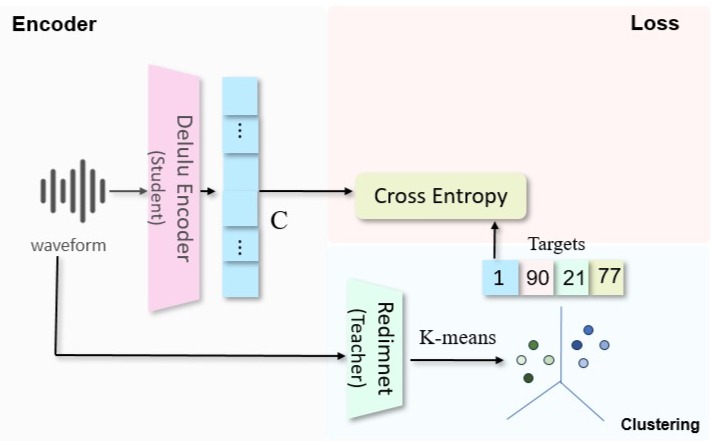} 
  \caption{Illustration of \textsc{DELULU} architecture. The student encoder is trained with a masked prediction objective, where frame-level features from the teacher (ReDimNet) are clustered into discrete speaker-aware targets for cross-entropy training.}

  \label{fig:delulufig}
\end{figure}

Based on the architecture depicted in Figure~\ref{fig:delulufig}, DELULU adopts a masked training design that integrates self-supervised learning with speaker-discriminative guidance through an external teacher model.
\subsection{DELULU Encoder}
The DELULU encoder follows the wav2vec~2.0 architecture~\cite{baevski2020wav2vec}, consisting of a convolutional feature extractor, a Transformer encoder, a projection layer, and a code embedding layer. 
The convolutional encoder is composed of seven 512-channel layers with strides \([4, 2, 2, 2, 2, 2, 2]\) and kernel widths \([10, 3, 3, 3, 3, 2, 2]\), differing from HuBERT~\cite{hsu2021hubert} by adjusting the stride pattern to ensure temporal alignment with the teacher model’s frame-level outputs. 
For 16 kHz input speech, this produces a latent feature sequence with a 16 ms frame rate (256$\times$ down-sampling factor). 
These latent features are then passed through a stack of Transformer blocks that model long-range dependencies, after which a projection layer reduces dimensionality and prepares features for clustering and loss computation. 
Overall, the DELULU encoder serves as the \textbf{student model} that learns to predict clustered representations.  






\subsection{Teacher-Guided Clustering}
Unlike conventional self-supervised speech models that rely purely on acoustic clustering, DELULU leverages \textbf{ReDimNet}~\cite{yakovlev2024reshape}, a state-of-the-art speaker verification model, to guide the pseudo-label generation process. 
Instead of using pooled utterance-level embeddings, we extract \textit{prepooled frame-level features} from ReDimNet to preserve temporal resolution and ensure alignment with DELULU’s encoder outputs. 
These features are then used for k-means clustering with $k=256$ clusters. 
This approach ensures that the discrete targets encode speaker-specific characteristics such as voice quality, prosody, and speaking style, rather than prioritizing only phonetic content. 

\subsection{Training}
DELULU is trained using a masked prediction loss.  

\paragraph{Masked Prediction Loss:} Following the masked language modeling paradigm, a portion of the input time steps are randomly masked. The model is trained to predict the cluster assignments (derived from ReDimNet-guided k-means) at the masked positions. Given the model's output $\mathbf{C}_i$ at masked position $i$ and the corresponding target cluster ID, the cross-entropy loss is computed as:
\begin{equation}
    \mathcal{L}_{\text{mask}} = -\sum_{i \in \mathcal{M}} \log P(c_i^* \mid \mathbf{C}_i)
\end{equation}
where $\mathcal{M}$ denotes the set of masked positions, $\mathbf{C}_i$ represents the logits over the $k=256$ possible clusters, and $c_i^*$ is the ground-truth cluster assignment obtained from the teacher model.


\section{Experimental Setup}
We build our system on top of the official \texttt{torchaudio} self-supervised learning recipes. 
For clustering, we adopt the MiniBatchKMeans algorithm from \texttt{scikit-learn}, with a mini-batch size of 10,000 frames. 
Initialization is performed with k-means++ using 20 random starts to ensure stability, and cluster assignments are set to $k=256$.  

Pre-training is conducted on 960 hours of LibriSpeech audio using 4 NVIDIA H100 GPUs, with each GPU processing 87.5 seconds of audio per batch. 
The model is trained for a total of 400k updates. 
Optimization follows the AdamW optimizer ($\beta_1=0.9, \beta_2=0.98$) with an initial learning rate of 5e-4. 
A linear warmup is applied for the first 32k steps, followed by a polynomial decay back to zero. 
To regularize training, we apply a weight decay of 0.01 and clip gradients at a maximum norm of 10.0.  

\begin{table*}[t]
\centering
\label{tab:ablation}
\begin{tabular}{lcccc}
\toprule
\textbf{Features} & \textbf{Clusters} & \textbf{Stride} & \textbf{EER (\%)} & \textbf{Rel. Impr.} \\
\midrule
MFCC & 100 & 20 & 37.73 & - \\
HuBERT (Stage 2) & 500 & 20 & 34.05 & 9.8\% \\
\midrule
ReDimNet & 256 & 16 & \textbf{13.53} & \textbf{60.2\%} \\
ReDimNet & 500 & 16 & 14.16 & 58.4\% \\
ReDimNet & 1024 & 16 & 14.16 & 58.4\% \\
ReDimNet & 256 & 15 & 14.16 & 58.4\% \\
\bottomrule
\end{tabular}
\caption{Ablation study on VoxCeleb1-O upstream speaker verification (EER \%). Lower is better. Results demonstrate that ReDimNet-guided clustering is the primary factor in DELULU's performance, with optimal cluster size k=256 and stride=16 for temporal alignment.}
\end{table*}

\begin{table}[t]
\centering
\resizebox{\columnwidth}{!}{  
\begin{tabular}{lccc}
\toprule
\textbf{Method} & \textbf{Vox1-O} & \textbf{LibriSpeech} & \textbf{Avg.} \\
\midrule
DELULU-Utt   & 34.62 & 11.18 & 22.90 \\
DELULU-Frame & \textbf{13.53} & 15.95 & \textbf{14.74} \\
\midrule
\textit{Rel. Impr.} & \textit{60.9\%} & \textit{-42.7\%} & \textit{35.6\%} \\
\bottomrule
\end{tabular}
}
\caption{Ablation study comparing frame-level versus utterance-level pseudo-labeling (EER \%). Lower is better. DELULU-Frame demonstrates superior cross-domain generalization on VoxO1 despite comparable in-domain performance, validating the benefit of fine-grained temporal supervision.}
\label{tab:frame_vs_utt}
\end{table}



\begin{figure}[t]
  \centering
  \includegraphics[width=0.99\linewidth]{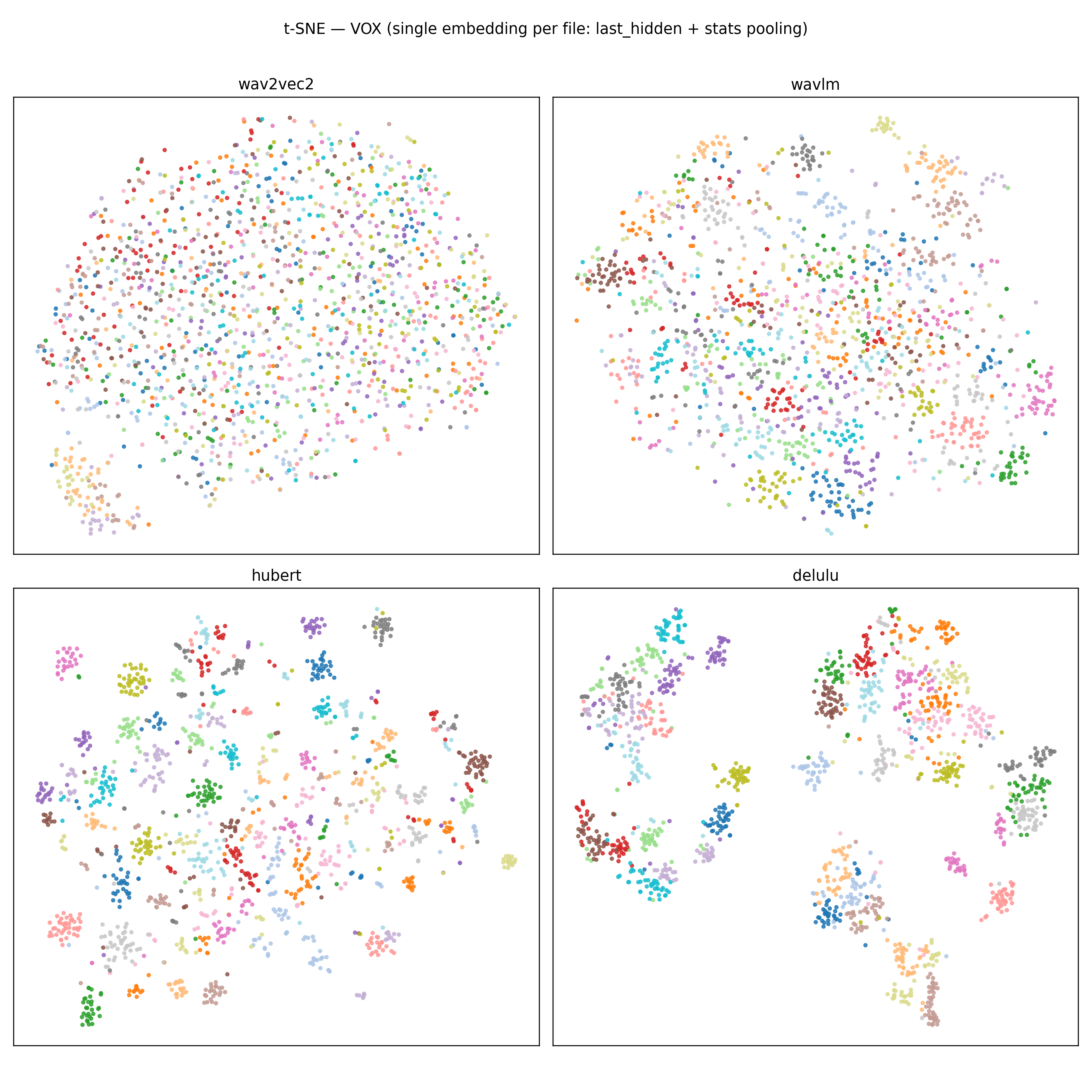}
    \caption{t-SNE visualization of embeddings from 40 VoxO1 speakers. \textsc{DELULU} yields compact, well-separated clusters with lower intra-speaker variability and higher inter-speaker separation than baseline SSL models.}
  \label{fig:tsneVox}
\end{figure}

\begin{figure}[t]
  \centering
  \includegraphics[width=0.99\linewidth]{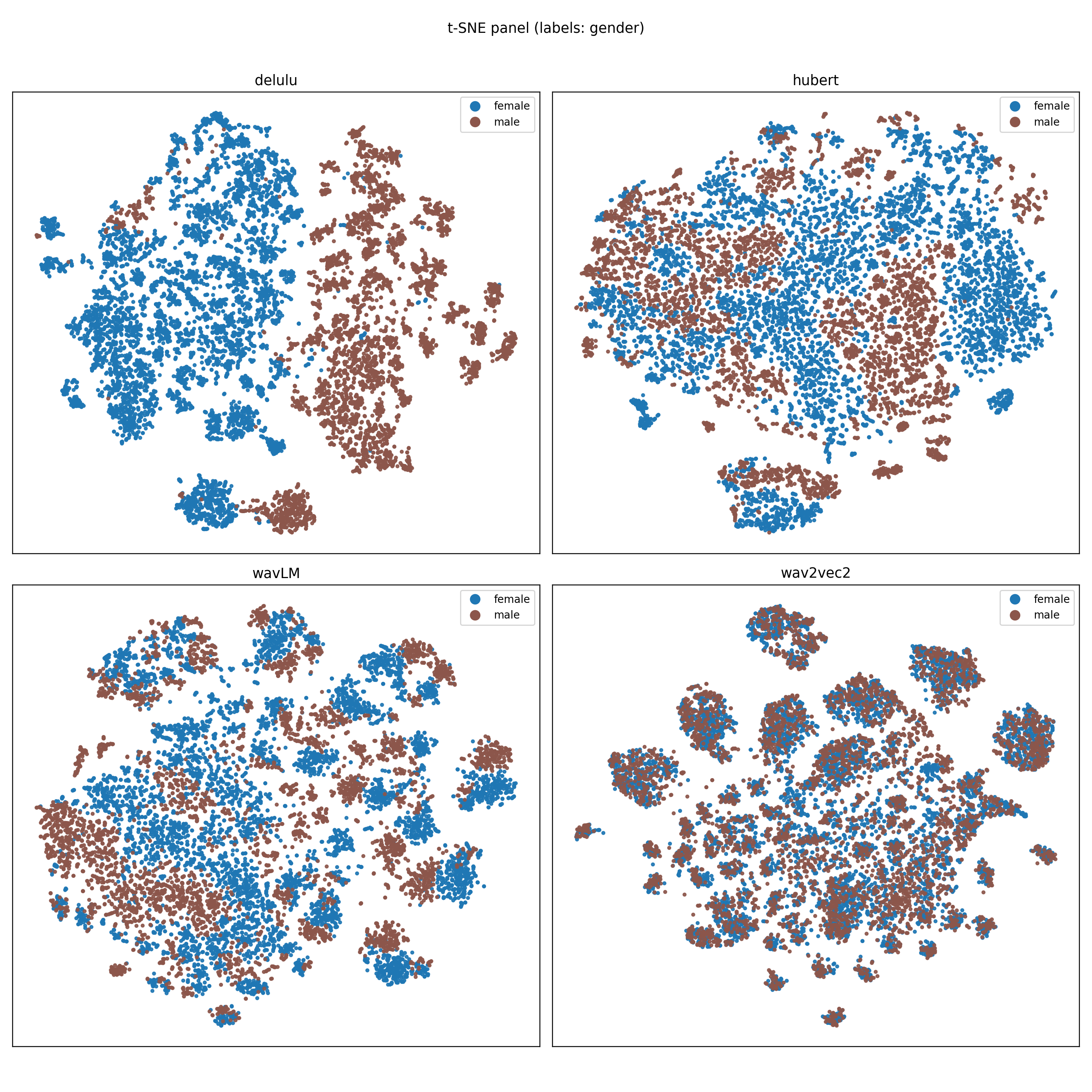}
  \caption{t-SNE visualization of speaker embeddings by gender on the EARS dataset. \textsc{DELULU} shows clear speaker clusters and strong gender separation.}
  
  \label{fig:tsneSveritasGender}
\end{figure}

\begin{figure}[t]
  \centering
  \includegraphics[width=0.89\linewidth]{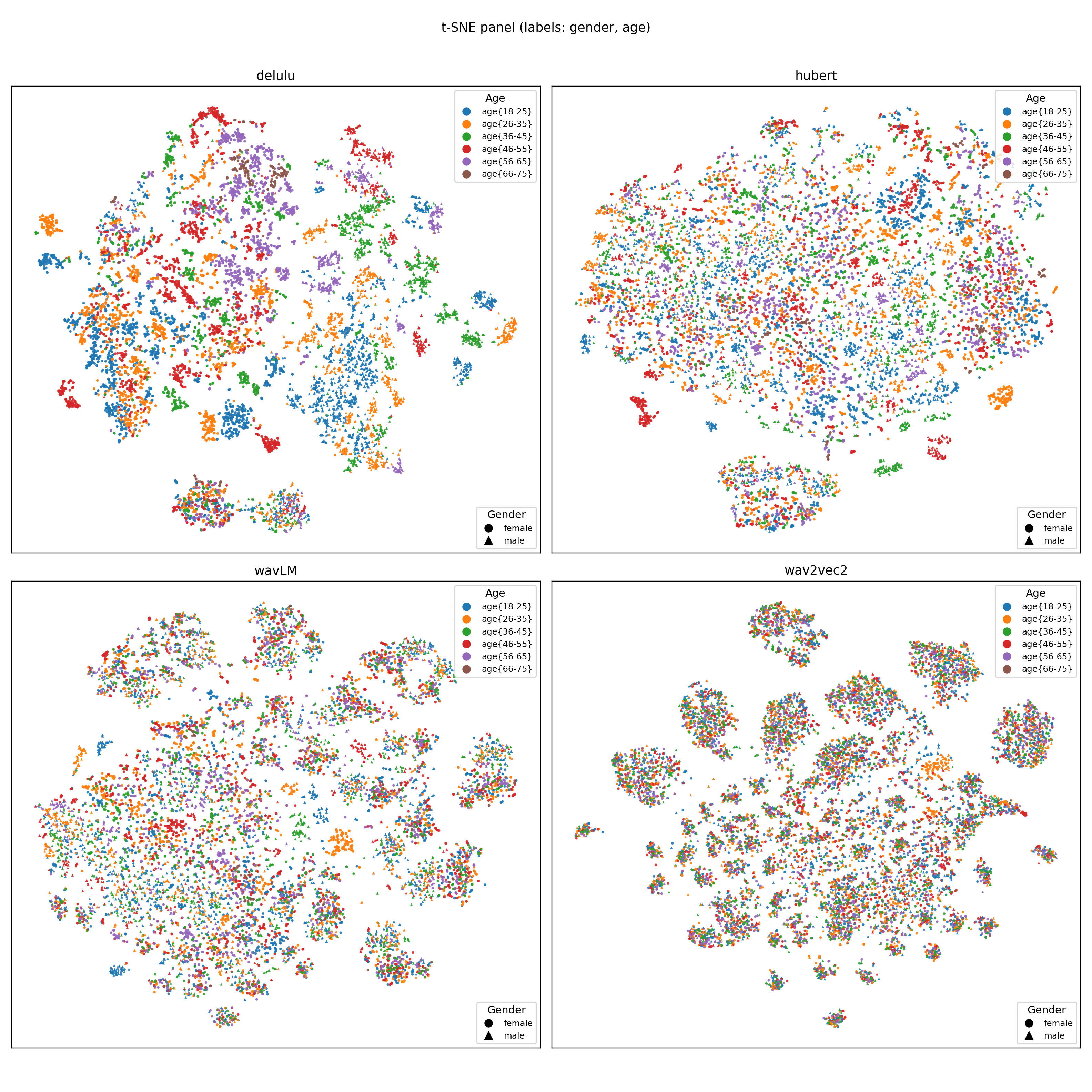}
  \caption{t-SNE visualization of EARS embeddings by gender and age groups (Table~\ref{tab:demographic_eer}). \textsc{DELULU} clusters primarily by speaker identity, with demographic traits forming secondary structure and no apparent bias.}

  \label{fig:tsneSveritasAgeGender}
\end{figure}

\section{Ablation Study}
To quantify the contribution of each component in DELULU's design, we perform systematic ablation experiments on the VoxCeleb1-O dataset, evaluating upstream speaker verification performance using EER, with all models trained for 400k updates on LibriSpeech 960h under identical optimization settings for fair comparison. Examining clustering feature sources, MFCC features with $k=100$ yield a baseline EER of 37.73\%, while higher-level acoustic representations from a pretrained HuBERT model with $k=500$ improve it to 34.05\%, indicating modest gains from context-aware features; in contrast, frame-level ReDimNet embeddings (dimension 2304) with $k=256$ reduce EER to 13.53\%, a 60\% relative improvement over HuBERT, confirming that external speaker-discriminative supervision during clustering drives DELULU's performance. For cluster size using ReDimNet features, we evaluate $k \in {256, 500, 1024}$, obtaining EERs of 13.53\%, 14.16\%, and 14.16\% respectively, showing $k=256$ offers the optimal trade-off between granularity and generalization, as 256 discrete speaker-aware units suffice for identity variation at this scale.
We also varied the DELULU encoder's stride, to find the optimal stride. Since the training paradigm requires frame-level synchrony between the teacher and student, the RedimNet stride was modified to be identical to that of the encoder. A stride of 16ms achieves 13.52\% EER, while both lower (15ms or less) and higher (20ms or greater) strides resulted in EERs greater than 14\%, hence we chose 16ms as our final stride.
Overall, these ablations highlight speaker-discriminative clustering as the key factor, with ReDimNet-guided pseudo-labels providing a 60\% relative improvement over HuBERT's acoustic-only approach, optimal $k=256$ balancing discriminability and generalization, and stride=16 ensuring effective supervision, validating the introduction of external speaker supervision into pseudo-label generation as a powerful strategy for speaker-aware representations. 

A central contribution of DELULU is the leveraging the power of frame-level pseudo-labeling, which enables masked prediction during pre-training and provides fine-grained temporal supervision. To quantify this benefit, we compare against an utterance-level variant (DELULU-Utt) where DELULU features are pooled, with pseudo-labels derived from ReDimNet utterance embeddings rather than frame-level representations. As shown in Table~\ref{tab:frame_vs_utt}, while DELULU-Utt achieves competitive in-domain performance on LibriSpeech (11.18\% EER), it suffers a substantial degradation on the out-of-domain VoxCeleb1-O benchmark (34.62\% EER). In contrast, DELULU-Frame maintains consistent performance across both evaluations (13.53\% and 15.95\% EER), yielding a 35.6\% relative improvement in average EER. These results demonstrates that frame-level supervision not only facilitates masked prediction but also promotes more robust speaker representations that generalize beyond the training distribution. Notably, while utterance-level clustering aligns pseudo-labels directly with the downstream verification objective—where decisions are inherently made at the utterance level—this apparent advantage leads to overfitting on in-domain data. By contrast, frame-level supervision forces the model to learn temporally localized speaker characteristics, yielding representations that transfer more effectively to unseen acoustic conditions.

\section{Evaluation}
\begin{table}[t]
\centering
\resizebox{\columnwidth}{!}{  
\begin{tabular}{llccc}
\hline
\textbf{Dataset} & \textbf{WavLM} & \textbf{Hubert} & \textbf{DELULU} & \textbf{Wav2Vec2} \\
\hline
VoxO1 & 35.93 &  34.05 & \textbf{13.53} &  43.17 \\
SITW & 44.00 & 42.60 & \textbf{25.40} & 42.20  \\
\hline
\end{tabular}
}
\caption{\label{tab:eer_sv} 
EER (\%) across upstream speaker verification benchmarks (↓ better)} 
\end{table}

\begin{table}[t]
\centering
\label{tab:ablation}
\resizebox{\columnwidth}{!}{%
\begin{tabular}{lcccc}
\toprule
\textbf{Features} & \textbf{Clusters} & \textbf{EER (\%)} & \textbf{Rel. Impr.} \\
\midrule
MFCC & 100 & 13.00 & -- \\
HuBERT (Stage 2) & 500 & 7.45 & -- \\
\midrule
ReDimNet & 256 & \textbf{5.63} & \textbf{24.5\%} \\
ReDimNet & 500 & 6.28 & 15.7\% \\
ReDimNet & 1024 & 5.99 & 19.6\% \\
\bottomrule
\end{tabular}
}
\caption{\label{tab:eer_sv_down} 
EER (\%) across downstream speaker verification (VoxO1) (↓ better)}
\end{table}
\begin{table*}[t]  
\centering 
\begin{tabular}{llccc}
\hline
\textbf{Task} & \textbf{WavLM} & \textbf{Hubert} & \textbf{DELULU} & \textbf{Wav2Vec2} \\
\hline
Spoof Detection & 51.44 &  53.51 & \textbf{57.20} &  52.88 \\
Speaker Count & 62.71 & 64.83 & \textbf{67.13} & 64.20  \\
Accent Detection & 77.76 & 62.86 & \textbf{78.38} & 58.60  \\
HowFarSpk & 69.37 & 70.58 & \textbf{73.36} & 57.84  \\
Gender Detection & 95.75 & 93.97 & \textbf{96.18} & 92.73  \\
Age Estimation & 32.69 & 29.43 & \textbf{36.00} & 31.99  \\

\hline
\end{tabular}
\caption{\label{tab:f1_knn_tasks} 
Zero-shot Macro-F1 score (\%) across profiling tasks (↑ better).} 
\end{table*}  
\begin{table}[t]
\centering
\caption{Teacher vs. Student comparison on zero-shot profiling tasks. We report accuracy (\%) at both frame-level and utterance-level evaluation. ReDimNet (teacher) was trained on VoxBlink2 ($>$100K speakers), while DELULU (student) was trained only on LibriSpeech ($\sim$1K speakers).}
\label{tab:frame_vs_utt}
\resizebox{\columnwidth}{!}{%
\begin{tabular}{llcc}
\toprule
\textbf{Evaluation} & \textbf{Task} & \textbf{ReDimNet} & \textbf{DELULU} \\
\midrule
\multirow{3}{*}{Frame-Level} 
 & Speaker Count & 47.00 & \textbf{51.00} \\
 & HowFarAreYou & 50.72 & \textbf{55.35} \\
 & Spoof Detection & 89.50 & \textbf{91.50} \\
\midrule
\multirow{3}{*}{Utterance-Level} 
 & Speaker Count & 29.50 & \textbf{65.50} \\
 & HowFarAreYou & 51.00 & \textbf{73.58} \\
 & Spoof Detection & 90.50 & \textbf{92.08} \\
\bottomrule
\end{tabular}%
}
\end{table}

We evaluate DELULU across multiple speaker-centric benchmarks described below to assess its effectiveness in capturing speaker-discriminative information. To verify our hypothesis that, as a more speaker-aware foundation model, DELULU may be expected to result in representations that are naturally better organized by speaker characteristics, and deliver better performance both in zero-shot ``upstream'' settings where they are used as-is, and in ``downstream'' settings where the model is further fine-tuned to the task, we perform the evaluations (comparing it to other baseline models) in both settings, with a primary focus on speaker verification for the downstream task. We also visualize the derived repersentations for further confirmation of our hypothesis. We report on these tests below.

For fair comparison, all baseline models (wav2vec 2.0, HuBERT, and WavLM) are pretrained on the same 960 hours of LibriSpeech data, ensuring that differences in performance arise from architectural and training objective choices rather than data scale.

\subsection{Upstream Evaluation: Speaker Verification}
We first evaluate the models in an upstream setting, where representations are extracted directly from the pretrained encoders without any fine-tuning. This zero-shot evaluation reveals the inherent speaker-discriminative properties encoded during self-supervised pretraining.

\paragraph{Protocol.} We evaluate on two widely-used speaker verification benchmarks: VoxO1~\cite{nagrani2017voxceleb} and SITW~\cite{mclaren2016speakers} (see Appendix~\ref{sec:appendix-sv}). For each model, we extract utterance-level embeddings by mean pooling over the temporal dimension of final layer representations. Verification trials are scored using cosine similarity, and we report EER, where lower is better.

\paragraph{Results.} Table~\ref{tab:eer_sv} shows that DELULU achieves substantial improvements over all baseline SSL models. On VoxO1, DELULU obtains an EER of 13.53\%, representing a 62\% relative improvement over HuBERT (34.05\%) and a 60\% improvement over WavLM (35.93\%). Similar gains are observed on SITW, where DELULU achieves 25.40\% EER compared to HuBERT's 42.60\% and WavLM's 44.00\%. These results demonstrate that ReDimNet guided clustering successfully embeds strong speaker-discriminative structure into the learned representations.





\subsection{Downstream Evaluation: Speaker Verification}
To assess transferability beyond upstream frozen representations, we evaluate \textbf{DELULU} on a downstream speaker verification task using VoxCeleb1-O. Since our objective is to evaluate DELULU's transferability rather than optimize for maximum downstream performance, we employ a minimal fine-tuning architecture that adds only a simple classification head on top of the frozen encoder, as described in Appendix \ref{sec:appendix-downstream-arch} and then compute EER to measure verification accuracy.  

As shown in Table~\ref{tab:eer_sv_down}, DELULU achieves 5.63\% EER with $k=256$ clusters, representing a 24.5\% relative improvement over HuBERT Stage 2 (7.45\%) and a 56.7\% gain over the MFCC baseline (13.00\%). 
Larger vocabularies ($k=500, 1024$) yield marginal improvements, confirming that $k=256$ provides the optimal balance between discriminability and generalization. 
These results demonstrate that the speaker-discriminative structure learned during pretraining transfers effectively to supervised fine-tuning, further validating DELULU as a strong foundational encoder for speaker verification.
\subsection{Demographic Subgroup Analysis}

To understand how speaker-discriminative information is distributed across demographic groups following the \textsc{SVeritas} \cite{baali2025sveritas} benchmark, we analyze model performance on the EARS \cite{richter2024ears} dataset (details in Appendix \ref{sec:appendix-ears-data}), stratified by gender and age subgroups.

\paragraph{Protocol.} Following the upstream protocol, we extract mean-pooled embeddings and compute EER for speaker verification trials within each demographic subgroup. This analysis reveals whether models learn speaker representations uniformly across age and gender categories or exhibit demographic biases.

\paragraph{Results.} Appendix \ref{sec:appendix-ears-data} Table~\ref{tab:demographic_eer} presents EER across gender and age subgroups. DELULU consistently outperforms baselines across all demographics, with particularly strong improvements for middle-aged speakers (36-55 years). For male speakers aged 36-45, DELULU achieves 24.53\% EER compared to HuBERT's 39.47\%. The consistent improvements across subgroups indicate that DELULU's speaker-discriminative representations generalize effectively across demographic variations without introducing systematic biases.

\subsection{Zero-Shot Speaker Profiling Tasks}
Beyond verification, we assess \textbf{DELULU}’s capacity to encode diverse speaker attributes through zero-shot evaluation on multiple profiling tasks from the DynamicSUPERB benchmark~\cite{huang2024dynamic}. 
We evaluate six speaker-related tasks, including age, gender, accent, speaker counting, and spoof detection (details in Appendix~\ref{sec:appendix-zeroshot-tasks}). 

\paragraph{Protocol.} 
We extract layer-wise representations from each model and train K-Nearest Neighbors (KNN) classifiers without any fine-tuning. 
All evaluations are conducted in a fully zero-shot setting; no supervised training is used for adaptation of the models. 
Given that these models are trained without explicit supervision, we expect performance to be poor, possibly approaching random levels; however, all SSL-based models achieve meaningful results, with \textsc{DELULU} consistently outperforming other models. 
For each task, we perform k-fold cross-validation and report Macro-F1 scores averaged across folds, along with standard deviations to indicate variability and robustness.

\paragraph{Results.} 
Table~\ref{tab:f1_knn_tasks} summarizes the zero-shot results across six profiling tasks. 
\textsc{DELULU} achieves the highest performance on all tasks, considerably outperforming other SSL models in zero-shot settings. 
The second-best model varies by task—for instance, WavLM performs competitively on gender classification, while HuBERT fares better on accent detection. 
On spoof detection, DELULU achieves 57.20\% F1 compared to HuBERT’s 53.51\% and WavLM’s 51.44\%; on accent detection, it reaches 78.38\% F1, surpassing HuBERT (62.86\%) and wav2vec 2.0 (58.60\%). 
Even on more challenging tasks such as age estimation (36.00\% F1), DELULU exhibits clear gains despite the absence of task supervision. 
These findings demonstrate that the speaker-guided clustering in DELULU enables robust and discriminative representations that generalize across diverse zero-shot profiling scenarios.

\subsection{Teacher vs. Student Evaluation.}
To evaluate the robustness of \textsc{DELULU}'s representations under different evaluation protocols, we compare frame-level and utterance-level performance against its teacher model, ReDimNet, on three profiling tasks: speaker counting, speaker distance estimation (HowFarAreYou), and spoof detection. For utterance-level evaluation, we apply mean pooling over frame embeddings, while for frame-level evaluation, we classify each frame independently using KNN and aggregate predictions via majority voting. Notably, ReDimNet was trained on VoxBlink2~\cite{lin2024voxblink2}, a large-scale audio-visual dataset comprising over 100,000 speakers with diverse acoustic conditions, whereas \textsc{DELULU} was trained solely on LibriSpeech~\cite{panayotov2015librispeech}, a significantly smaller corpus of approximately 1,000 speakers. Despite this substantial difference in training data scale and diversity, Table~\ref{tab:frame_vs_utt} shows that \textsc{DELULU} consistently outperforms ReDimNet on all three tasks under both evaluation protocols. On speaker counting, \textsc{DELULU} achieves 51.00\% and 65.50\% accuracy at the frame and utterance levels respectively, compared to ReDimNet's 47.00\% and 29.50\% Similarly, on HowFarAreYou, \textsc{DELULU} reaches 55.35\% (frame) and 73.58\% (utterance) versus ReDimNet's 50.72\% and 51.00\% For spoof detection, \textsc{DELULU} maintains a consistent edge at both frame level (91.50\% vs. 89.50\%) and utterance level (92.08\% vs. 90.50\%). These results demonstrate that \textsc{DELULU}'s speaker-guided distillation enables superior generalization to zero-shot profiling tasks, even when trained on way smaller data compared to its teacher.

\subsection{Representation Analysis}

We analyze the learned embedding space using t-SNE visualizations to assess whether DELULU produces structured and discriminative speaker representations.  
In Figure~\ref{fig:tsneVox}, we expect embeddings from the same speaker to form compact clusters, distinct from those of other speakers. Indeed, DELULU’s representations exhibit clear and well-separated speaker clusters, indicating strong identity preservation and reduced intra-speaker variability compared to baseline models.  
In Figure~\ref{fig:tsneSveritasGender}, we expect representations to separate naturally by gender. The visualization confirms this behavior, showing distinct groupings corresponding to male and female speakers, while maintaining tight clustering within each group.  
Finally, in Figure~\ref{fig:tsneSveritasAgeGender}, we examine whether age-related patterns emerge within the embedding space. As expected, DELULU produces smooth transitions across age groups, suggesting that the model implicitly encodes demographic cues such as vocal maturity and pitch characteristics alongside dominant speaker identity.  
These results demonstrate that DELULU learns a representation space that reflects both speaker individuality and meaningful demographic structure, confirming the model’s capacity for fine-grained, interpretable speaker encoding.



\section{Conclusion}

We introduce \textsc{DELULU}, a speaker-aware self-trained foundational model that enhances speaker representation learning by guiding pseudo-label generation with frame-level embeddings from ReDimNet. Across benchmarks in speaker verification, profiling, and counting, DELULU consistently outperforms existing SSL models, establishing it as a strong universal encoder for speaker-centric tasks. Since the core idea lies in incorporating task-relevant structure into pseudo-label generation without direct supervision, we expect this strategy to generalize to other foundational model architectures and downstream applications. Future work will explore replacing clustering with distillation-based objectives to further simplify the pretraining pipeline.

\section*{Limitations}
While \textsc{DELULU} demonstrates substantial improvements in speaker verification, profiling, and zero-shot speaker-aware tasks, several limitations warrant further investigation. 

Extending DELULU to large-scale, multi-domain datasets with higher speaker and environmental variability remains an open challenge. 
Also, incorporating an external model (ReDimNet) into the clustering process introduces additional computational overhead and memory requirements, potentially limiting accessibility for research groups with restricted computational resources.  
Finally, while DELULU excels in identity-aware modeling, its downstream adaptability to non-speaker-centric applications, such as emotion or intent recognition, has yet to be comprehensively explored. 
Future work will focus on addressing these limitations through more efficient training strategies, broader data coverage, and cross-domain generalization studies.

\section*{Ethics Statement}
The development of \textsc{DELULU} is guided by a strong commitment to ethical responsibility and privacy preservation. 
As a speaker-aware foundational model, DELULU possesses the potential to be misused for surveillance, impersonation, or identity inference without consent. 
We emphasize that its use must strictly comply with ethical research standards, data protection laws, and institutional review protocols. 
All datasets employed in this study (LibriSpeech, VoxCeleb, DynamicSuperb, and SVeritas) consist of publicly available and consented speech samples intended solely for research purposes. 
No personally identifiable or private data was used or generated. 
Furthermore, while DELULU improves representation learning for speaker-related tasks, it does not perform speaker identification on unconsented audio, nor is it intended for forensic or monitoring applications. 
In line with the ACL Ethics Policy, we advocate for transparent deployment of DELULU, ensuring that its advancements contribute positively to responsible and equitable speech technology research.

\bibliography{main}

\appendix
\newpage
\section{Experimental Setup}
\paragraph{Computational complexity} 
Compared to the HuBERT baseline, DELULU introduces additional computational cost during the preprocessing stage due to RedimNet’s convolutional feature extraction. However, this overhead is performed offline and does not affect training or inference. Empirically, on an NVIDIA H100 GPU, HuBERT requires approximately 14.5\,ms per batch during inference, while RedimNet requires approximately 51.1\,ms under identical hardware and batch conditions. Since preprocessing is handled once and cached, this difference does not impact overall training efficiency or deployment latency in our setup.

\section{Upstream Evaluation}
\paragraph{SV Benchmarks} 
\label{sec:appendix-sv}

\begin{itemize} [leftmargin=*]
    \item \textbf{VoxCeleb1-O}~\cite{nagrani2017voxceleb}: The original VoxCeleb1 test set containing 37,720 verification trials across 40 speakers.
    \item \textbf{SITW}~\cite{mclaren2016speakers}: Speakers in the Wild, a challenging dataset with 6,445 trials featuring diverse acoustic conditions and speaking styles.
    \textbf{LibriSpeech} 
\end{itemize}

\section{Downstream Model}
\label{sec:appendix-downstream-arch}

\paragraph{Model Architecture.} Our downstream model consists of the pretrained SSL encoder (frozen) followed by temporal mean pooling and an L2 normalization layer. A single linear embedding head projects the pooled representations to speaker logits for classification. This minimal design isolates the quality of the pretrained representations by limiting the capacity of the downstream classifier.

\paragraph{Training Setup.} We train the embedding head on VoxCeleb development set using cross-entropy loss with the AM-Softmax objective~\cite{wang2018additive}. The SSL encoders remain frozen throughout training, ensuring that performance differences reflect the quality of pretrained features rather than fine-tuning capacity. We train for 30 epochs with a batch size of 32 and learning rate of 1e-3.

\section{Demographic Subgroup Analysis}
\label{sec:appendix-ears-data}
\paragraph{Description of the EARS Dataset.} The EARS (Expressive Anechoic Recordings of Speech) dataset contains speakers spanning ages 18-75, with balanced gender representation. We evaluate on 102 speakers (59 female, 43 male) across six age brackets.
\begin{table*}[t]
\centering
\begin{tabular}{llcccccc}
\hline
\textbf{Category} & \textbf{Subgroup} & \textbf{WavLM} & \textbf{Hubert} & \textbf{DELULU} & \textbf{Wav2Vec2} \\
\hline
\multirow{2}{*}{Gender} 
& Female (59 spks) & 41.69 & 41.06 & \textbf{28.11} & 45.60 \\
& Male (43 spks)   & 40.76 & 41.47 & \textbf{28.54} & 44.70 \\
\hline
\multirow{11}{*}{Age} 
& F (18–25), 13 spks & 41.93 & 39.67 & \textbf{31.01} & 45.03 \\
& F (26–35), 13 spks & 40.60 & 41.61 & \textbf{30.90} & 43.90 \\
& F (36–45), 7 spks  & 43.37 & 44.26  & \textbf{29.38} & 47.40 \\
& F (46–55), 14 spks & 42.17 & 40.48 & \textbf{29.56} & 45.78 \\
& F (56–65), 10 spks & 42.71 & 42.02 & \textbf{31.28} & 47.78 \\
& F (66–75), 2 spks  & 43.66 & 42.25 & \textbf{40.12} & 48.32 \\
& M (18–25), 14 spks & 41.55 & 43.06 & \textbf{35.01} & 45.87 \\
& M (26–35), 10 spks & 40.51 & 40.60 & \textbf{30.90} &  42.97 \\
& M (36–45), 10 spks & 39.40 & 39.47 & \textbf{24.53} & 45.35 \\
& M (46–55), 4 spks  & 41.27 & 42.55 & \textbf{29.63} & 44.25 \\
& M (56–65), 5 spks  & 42.53 & 43.32 & \textbf{31.37} & 47.53 \\
\hline
\end{tabular}
\caption{\label{tab:demographic_eer}EER (\%) across demographic subgroups in EARS dataset on upstream speaker verficiation (↓ better)}
\end{table*}

\section{Zero-Shot}
\paragraph{Profiling Tasks.} 
\label{sec:appendix-zeroshot-tasks}

\begin{itemize}
    \item \textbf{Spoof Detection}: Binary classification task detecting synthesized or manipulated speech versus genuine human speech.
    \item \textbf{Speaker Counting}: Predicting the number of unique speakers in an audio segment.
    \item \textbf{Accent Detection}: Classifying the regional accent of the speaker.The dataset includes nine distinct distance classes.
    \item \textbf{HowFarSpk}: Determining the spatial distance between speaker and microphone. The dataset includes three distinct distance classes. 
    \item \textbf{Gender Recognition}: Binary classification of speaker gender.
    \item \textbf{Age Classification}: Multi-class prediction of speaker age group.
\end{itemize}

\label{sec:appendix}

\end{document}